\documentclass{article}

\usepackage{PRIMEarxiv}

\usepackage[utf8]{inputenc} 
\usepackage[T1]{fontenc}    
\usepackage{hyperref}       
\usepackage{url}            
\usepackage{booktabs}       
\usepackage{amsfonts}       
\usepackage{nicefrac}       
\usepackage{microtype}      
\usepackage{lipsum}
\usepackage{fancyhdr}       
\usepackage{graphicx}       
\usepackage{multicol}
\usepackage{multirow}
\usepackage{nccmath}

\graphicspath{{media/}}     

\title{pinnet: a deep neural network with pathway prior knowledge for Alzheimer's disease
}

\author{
  Yeojin Kim \\
  Artificial Intelligence Graduate School \\
  Gwangju Institute of Science and Technology\\
  Gwangju, South Korea \\
  \texttt{hapy0907@gm.gist.ac.kr} \\
   \And
  Hyunju Lee * \\
  School of Electrical Engineering and Computer Science\\
  Gwangju Institute of Science and Technology\\
  Gwangju, South Korea \\
  \texttt{hyunjulee@gist.ac.kr} \\
}

\begin{document}
\maketitle

\begin{abstract}
Identification of Alzheimer’s Disease (AD)-related transcriptomic signatures from blood is important for early diagnosis of the disease. Deep learning techniques are potent classifiers for AD diagnosis, but most have been unable to identify biomarkers because of their lack of interpretability. To address these challenges, we propose a pathway information-based neural network (PINNet) to predict AD patients and analyze blood and brain transcriptomic signatures using an interpretable deep learning model. PINNet is a deep neural network (DNN) model with pathway prior knowledge from either the Gene Ontology or Kyoto Encyclopedia of Genes and Genomes databases. Then, a backpropagation-based model interpretation method was applied to reveal essential pathways and genes for predicting AD. We compared the performance of PINNet with a DNN model without a pathway. Performances of PINNet outperformed or were similar to those of DNN without a pathway using blood and brain gene expressions, respectively. Moreover, PINNet considers more AD-related genes as essential features than DNN without a pathway in the learning process. Pathway analysis of protein-protein interaction modules of highly contributed genes showed that AD-related genes in blood were enriched with cell migration, PI3K-Akt, MAPK signaling, and apoptosis in blood. The pathways enriched in the brain module included cell migration, PI3K-Akt, MAPK signaling, apoptosis, protein ubiquitination, and t-cell activation. Collectively, with prior knowledge about pathways, PINNet reveals essential pathways related to AD. 
\end{abstract}


\section{Introduction}
Alzheimer’s disease (AD) is the most prevalent type of dementia and is  distinguished by amyloid beta (A$\beta$) plaques and neurofibrillary tangles in the brain. A$\beta$ and tau are clinical hallmarks of AD, in which abnormal accumulation precedes neurodegeneration and cognitive impairment in both sporadic and familial AD \cite{bateman2012clinical}. Although much progress has been made in understanding AD pathology, currently available treatments modify only symptoms, such as cognitive and behavioral dysfunction \cite{yiannopoulou2020current}. The use of biomarkers for identifiying AD in the pre-dementia phase, prior to the appearance of clinical symptoms, is essential for the AD diagnosis\cite{huynh2017alzheimer}. Recently, AD has been increasingly recognized as a systemic disease, supported by numerous studies that have uncovered a peripheral mechanism for AD progression \cite{zhang2013systemic, morris2014alzheimer}. Brain-derived A$\beta$ can be cleared in the periphery by monocytes, which is boosted by the immune system\cite{cheng2020peripheral}. Multiple peripheral inflammatory markers were elevated in AD patients\cite{lai2017peripheral}, and it was showed that inflammation also plays an essential role in AD development. Furthermore, Urayama et al. \cite{urayama2022preventive} reported a significant reduction of A$\beta$ plaque development by 40-80\% and improvement in memory performance through exchanging the whole blood in AD mice with normal blood. However, the exact mechanism of how blood exchanges reduce amyloid pathology remains unclear.

Several studies based on gene expression data have been conducted to find biomarkers for AD. Puthiyedth \emph{et al.}\cite{puthiyedth2016identification} analyzed microarray gene expression data of 161 post-mortem brain samples across six brain regions and identified new AD candidate genes and 23 non-coding features. Xu \emph{et al.}\cite{xu2018systematic} analyzed gene expressions from brain tissues by constructing a transcriptomic network and demonstrated that activation of 17 hub genes, including YAP1, at the early stage could promote AD. Although gene expression data of brain tissues obtained from post-mortem autopsy revealed important molecular mechanisms regarding AD, their clinical application is limited due to their invasiveness. Instead, studies based on blood gene expression data have been conducted for early diagnosis of Alzheimer's disease. Li \emph{et al.}\cite{li2017identification} detected leukocyte-specific expression changes in peripheral whole blood gene expression data. They found that differentially expressed genes were associated with Wnt signaling pathways and mitochondrial dysfunction, suggesting a significant overlap in brain area expression profiles. However, few studies have compared the transcriptomic signatures of AD between blood samples and brain samples.

Many deep learning methods have been applied to biomedical domains and showed improved performances for various problems such as predictions of properties of DNA sequences, protein structure prediction, and disease classification \cite{luo2019deepphos, senior2020improved, mostavi2020convolutional}. In these methods, weights in neural networks are optimized for a given objective function, but they do not necessarily reveal a group of neurons and input features related to a given biomedical problem. Recently, several studies have started searching for a model to obtain outcomes with explainable mechanisms \cite{kuenzi2020predicting, lee2020cancer}. If a deep learning model that can predict samples' disease status and analyze the internal relationships between genes and diseases is developed, it might be used as both a classifier and a tool for finding biomarkers.

This study proposes a pathway information-based deep neural network (PINNet) to predict AD using a gene expression dataset from the brain and blood. Two sources of pathway information are employed: Gene Ontology (GO), a literature-curated reference database with a hierarchical structure, or the Kyoto Encyclopedia of Genes and Genomes (KEGG), an alternative pathway ontology containing dynamics and interaction between the genes. The GO ontology design is interrelated through a hierarchical parent-child relationship, enabling us to examine functional clusters at different scales. KEGG represents networks of interacting molecules responsible for specific biological functions. In PINNet, pathway information was included in the structure of the model, thereby enhancing interpretation. The model was interpreted using Deep SHAP \cite{lundberg2017unified}, which allows us to identify important predictor genes in the model. Thus, we evaluated the performance of PINNet models in blood and brain gene expression samples for two pathway databases. When we examined gene signatures of blood and brain that played essential roles in each predictive model,  contributing genes in PINNet were highly prominent in known sets of AD-related genes.

\section{Results}
\label{sec:headings}


\subsection{PINNet: A deep neural network with pathway prior knowledge for AD classification}
To aid in identifying transcriptional features regulating AD, we developed a deep neural network-based classification model named PINNet, which predicts AD based on pathway information. Figure 1 shows the structure of the PINNet. Contrary to previous deep learning approaches, PINNet does not rely on operating as a black box but instead focuses on prior biological knowledge from the GO or KEGG databases. PINNet consists of four layers: an input layer, a pathway layer, a fully connected hidden layer, and a softmax output layer, and the model is built to predict AD (see Fig. 1 and Methods). Blood or brain transcription datasets are fed into the input layer of the model, and the output layer indicats a sample status (AD or control[CN]). We formulated the structure of the model from genes to biological functional groups, enabling biological interpretation. The use of multiple levels of ontology (ranging from 10 to 1156 genes per pathway) acknowledges the diverse complexity of biological processes. During training, the expression data of each gene are induced to patterns of pathway-level of activities, enabling \emph{in silico} investigation of the biological processes underlying transcriptome-disease association. Genes are partially propagated through the same biological process that contains them, giving rise to functional changes at the pathway level, consequently predicting sample status. Using this design, the PINNet embedded in GO includes 4,026 biological process terms, and the corresponding model for KEGG includes 168 pathways. To incorporate the effect of the pathways, we partially masked the weights of the pathway layer using a binary matrix that represents relationships between pathways and genes. As these neurons' connection is designed to represent biological processes, pathway nodes only receive inputs from the genes included in the pathway. On average, one GO BP is connected to 84 input gene nodes and 42 input gene nodes for KEGG. We normalized pathway nodes with the size of pathways to ensure that the number of genes included in the pathway does not affect the importance of the pathway node. In addition, since the pathway node is very sparsely connected, there is a difference in the range of values of the pathway node and the fully connected node. Therefore, layer normalization was applied to the mini-batch when pathway nodes and fully connected nodes were concatenated. Afterward, they pass through one hidden layer and softmax to predict AD.

\begin{figure}
  \centering
  \includegraphics[width=\linewidth]{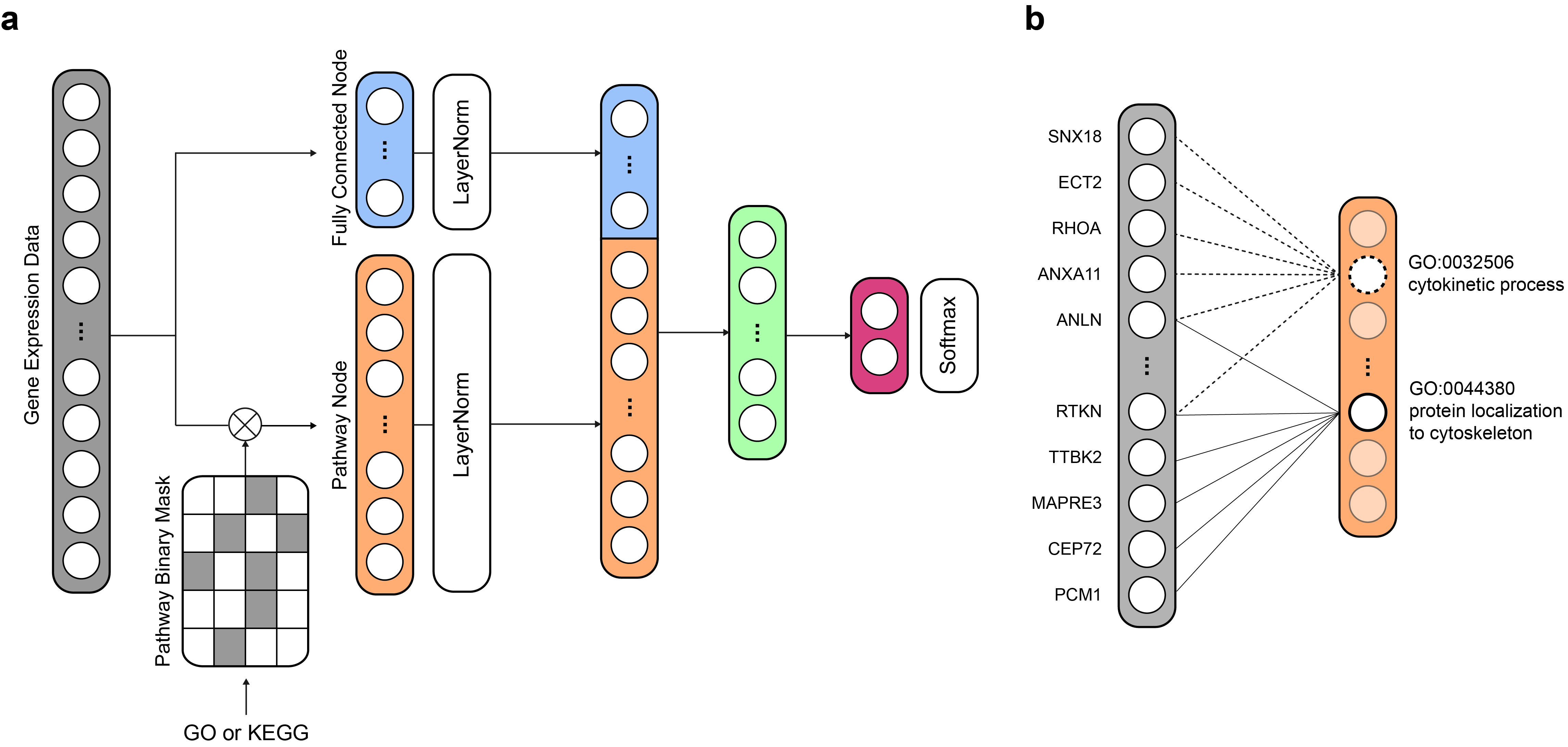}\par
  \caption{An overview of the model structure. (a) The model consists of an input layer, a pathway layer, a hidden layer, and a softmax layer. The mRNA expression dataset and pathway information were assigned to the input nodes and binary mask matrix, respectively. (b) The pathway layer connects the relationship between pathways and genes.}
  \label{fig:fig1}
\end{figure}

\subsection{Performance in AD prediction}
We applied PINNet to brain gene expression data \cite{narayanan2014common} and blood gene expression data from Alzheimer’s Disease Neuroimaging Initiative (ADNI) \cite{petersen2010alzheimer} and constructed four models using two pathway datasets: blood with GO [blood (GO)], blood with KEGG [blood (KEGG)], brain with GO [brain (GO)], and brain with KEGG [brain (KEGG)]. While constructing a sparse pathway network, we normalized the pathway nodes with the size of pathways. To identify an appropriate normalization method, we tried three types of a normalization vector $u$: (1) $u=1$ (without normalization), (2) $u = {1}/{n}$, and (3) $u = {1}/{\sqrt{n}}$, where $n$ is the number of genes in each pathway, and compared the importance score of pathway nodes (Details about an importance score are described in Methods). Without normalization, pathways with a larger number of genes had higher importance scores in all four models. The absolute values of the Pearson correlation coefficient between the numbers of genes in pathways and importance scores were strongly correlated in the blood (GO) ($|R|$ $=$ 0.43) and blood (KEGG) ($|R|$ $=$ 0.30). However, when normalizing with $\sqrt{n}$, $|R|$ $<$ 0.1 was obtained in four models. This implies that the number of genes in the pathway has almost no effect on the importance scores of the pathways through normalization. Based on this result, the value of the pathway node was normalized with $\sqrt{n}$. We trained PINNet to predict the disease status of samples (see Methods) and tested its performance by 10-fold-cross-validation. In the cross-validation, each fold preserved the ratio of sample classes, and the ratios of AD to CN were 0.46 and 1.97 in blood and brain, respectively. A part of the training set was used as a validation set so that the ratio of training, validation, and test set was 8:1:1. During the training of PINNet, a validation set was used for early stopping and selection of the number of fully connected nodes and learning rates with grid search. Early stopping was applied if the validation AUC did not increase by more than ten epochs during the training. The number of fully connected nodes in the pathway layer is a hyperparameter with a set of candidate values of 32, 64, and 128, smaller than the number of pathways. The number of nodes in the second hidden layer was fixed as 64. Candidate values for the learning rate are 0.0001, 0.0005, and 0.001. All compared methods were trained and tested on the same data splits.

With the brain gene expressions from the prefrontal cortex area \cite{narayanan2014common}, the area under the curve (AUC) values between true positives and false-positive rates were 0.9744 and 0.9763 with GO and KEGG pathways, respectively (Table 1). The previous study using the same dataset showed the accuracy of around 86.30\% to 91.22\%\cite{cheng2021machine} using several machine learning models, reporting that brain gene expressions are valuable to classifying AD patients. This was also confirmed in our study. With the blood gene expressions from ADNI, we obtained AUC values of 0.6355 and 0.6420 with GO and KEGG pathways, respectively (Table 1). Predictions using the blood gene expressions showed relatively lower classification performance than brain expression data. This range of performance was also reported in the previous study \cite{lee2020prediction}.

 We compared PINNet with SVM, random forest, and deep neural network (DNN) methods. The DNN model architecture is identical to PINNet with the same number of the first hidden layer, except that it does not apply a pathway binary mask. The DNN model consists of two fully connected layers without a pathway binary mask, and the number of nodes and learning rate of the first hidden layer were selected using grid search.  Since the range of nodes in the first hidden layer of PINNet was 200 (168+32) to 4154 (4026+128), the candidate nodes in the first hidden layer includes 128, 512, 1024, and 4096, and the learning rate used the same set of candidate values as PINNet. As shown in Table 1, our PINNet outperformed the SVM and RF models, achieving better classification performance on blood and brain datasets. Despite the extra parameters in DNN, which is a fully connected model, the performance of the DNN was lower than PINNet, especially for blood. Additionally, Table 2 shows F1 scores of PINNet and comparing models. 

\begin{table}[ht]
\centering
\begin{tabular}{|c|c|c|c|cc|}
\hline
\multirow{2}{*}{} & \multirow{2}{*}{SVM} & \multirow{2}{*}{RF} & \multirow{2}{*}{DNN} & \multicolumn{2}{c|}{PINNet}    \\ \cline{5-6} 
                  &                      &                     &                      & \multicolumn{1}{c|}{GO} & KEGG \\ \hline
Blood             & 0.5820$\pm$0.1073        & 0.5866$\pm$0.1112       & 0.6144$\pm$0.0789        & \multicolumn{1}{c|}{0.6355$\pm$0.1095}  & 0.6420$\pm$0.0811    \\ \hline
Brain             & 0.9536$\pm$0.0321        & 0.9717$\pm$0.0311       & 0.9744$\pm$0.0401        & \multicolumn{1}{c|}{0.9744$\pm$0.0329}  & 0.9763$\pm$0.0288    \\ \hline 
\end{tabular}

\caption{Comparison of performance (AUC) of different methods. Performance comparison for different machine learning and deep learning models. The DNN is a control model for each pathway dataset, consisting of a fully connected layer. SVM, support vector machine; RF, random forest; DNN, deep neural network; GO, gene ontology; AUC, area under the receiver operating characteristic curve.}
\label{tab:table1}
\end{table}

\begin{table}[ht]
\centering
\begin{tabular}{|c|c|c|c|cc|}
\hline
\multirow{2}{*}{} & \multirow{2}{*}{SVM} & \multirow{2}{*}{RF} & \multirow{2}{*}{DNN} & \multicolumn{2}{c|}{PINNet}    \\ \cline{5-6} 
                  &                      &                     &                      & \multicolumn{1}{c|}{GO} & KEGG \\ \hline
Blood             & 0.5448$\pm$0.0395        & 0.5405$\pm$0.0586       & 0.5563$\pm$0.0635        & \multicolumn{1}{c|}{0.5907$\pm$0.0762}  & 0.5542$\pm$0.0547    \\ \hline
Brain             & 0.9384$\pm$0.0265        & 0.9568$\pm$0.0339       & 0.9657$\pm$0.0327        & \multicolumn{1}{c|}{0.9607$\pm$0.0330}  & 0.9639$\pm$0.0321    \\ \hline
\end{tabular}

\caption{Comparison of performance (F1-score) of different methods. Performance comparison for different machine learning and deep learning models. The DNN is a control model for each pathway dataset, consisting of a fully connected layer.  SVM, support vector machine; RF, random forest; DNN, deep neural network; GO, gene ontology; AUC, area under the receiver operating characteristic curve.}
\label{tab:table2}
\end{table}

\begin{figure}
  \centering
  \includegraphics[width=\linewidth]{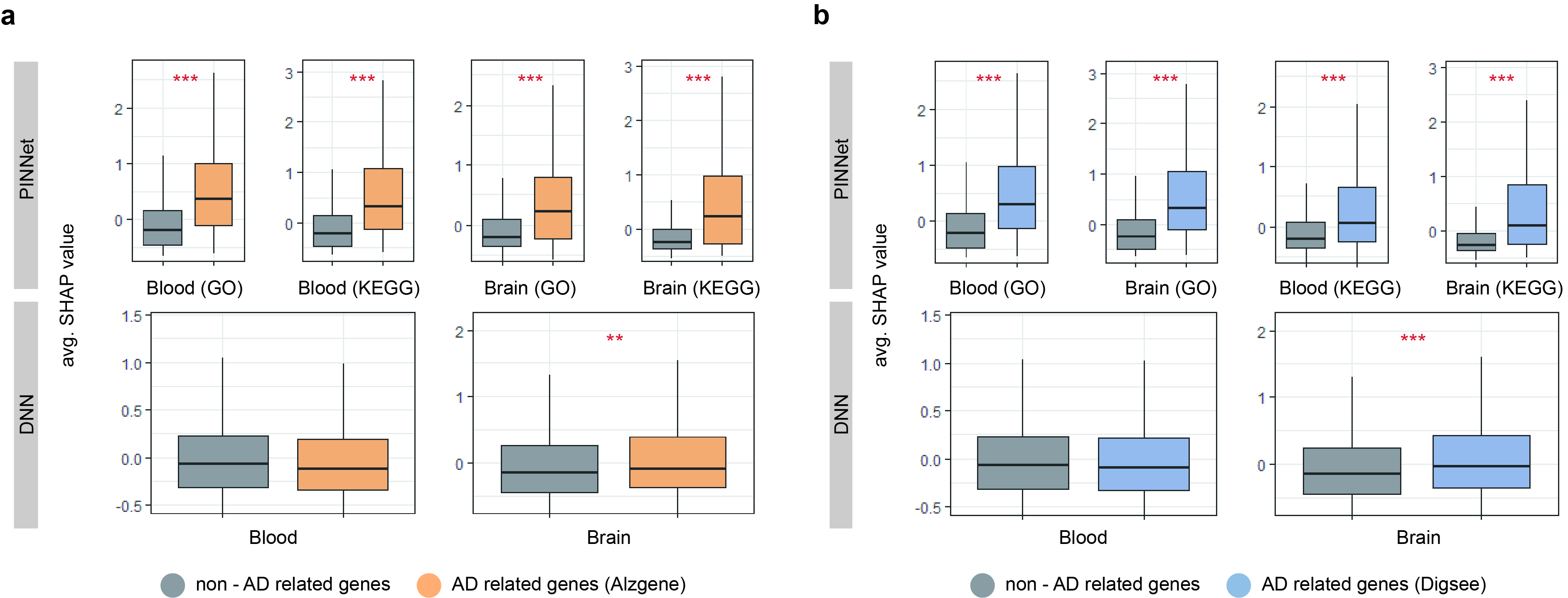}\par
  \caption{Importance score of known Alzheimer's disease (AD) genes in PINNet and DNN. We compared the SHAP (SHapley Additive exPlanations) values of known AD-related genes, including (a) Alzgene and (b) Digsee, in PINNet and DNN. The $y$-axes in each graph shows the average SHAP values. * denotes a $p$-value $\leq$ 0.05 of the Wilcoxon test between non-AD related genes and AD-related genes; ** indicates a $p$-value $\leq$ 0.01 of significance; *** represents a $p$-value $\leq$ 0.001.}
  \label{fig:fig2}
\end{figure}

\subsection{Known AD-related genes are important features of PINNet}
A major objective of the analysis of gene expressions for the disease is to associate changes in the transcript with changes in phenotype. To reveal such associations, we interpreted PINNet models for AD prediction. We identified which genes played an important role in prediction in the models learned from 10-fold-cross validation in Table 1. To determine the contribution of each gene in AD prediction, the feature importance of nodes in the input layer was measured by the SHAP (SHapley Additive exPlanations) value using the Deep SHAP method, a backpropagation-based deep neural network interpretation algorithm (see the Methods section). For each pathway (GO and KEGG) and each gene expression dataset (brain and blood), we calculated SHAP values from the trained PINNet models for each gene. We calculated the importance score as the average of the normalized SHAP values of the ten models. As a result, we obtained the importance scores of genes for each model of blood (GO), brain (GO), blood (KEGG), and brain (KEGG).

We analyzed the importance scores of genes using two lists of known AD-related genes: AlzGene \cite{bertram2007systematic} and DigSee \cite{kim2017analysis}. The AlzGene contains 681 genes curated by systematic meta-analyses, and 361 genes were shared with the 8,922 genes used in the study. DigSee contains 1,635 AD-related genes extracted from PubMed abstracts using text-mining techniques, and 961 were common with genes used in this study. We defined genes included in AlzGene or DigSee as AD-related genes and the remaining as non-AD-related genes. We used the Wilcoxon test to compare the importance score between AD-related and non-AD-related genes in PINNet and DNN. As shown in Figure 2, AD-related genes screened from DigSee and AlzGene showed significantly higher contributions than non-AD-related genes in PINNet models for all datasets ($p$-value $\le$ 0.001). However, in the case of DNN models, the difference in importance score between AD-related genes and non-AD-related genes is only partially significant for brain data, and there is no difference in blood. In addition, the difference in importance scores was larger in PINNet in all datasets. Unlike standard DNNs, PINNet captures important biological features based on prior biological knowledge. It represents that PINNet's biological prior knowledge leads to learning disease-related genes as essential features.

\begin{figure}
\centering
\includegraphics[width=0.9\linewidth]{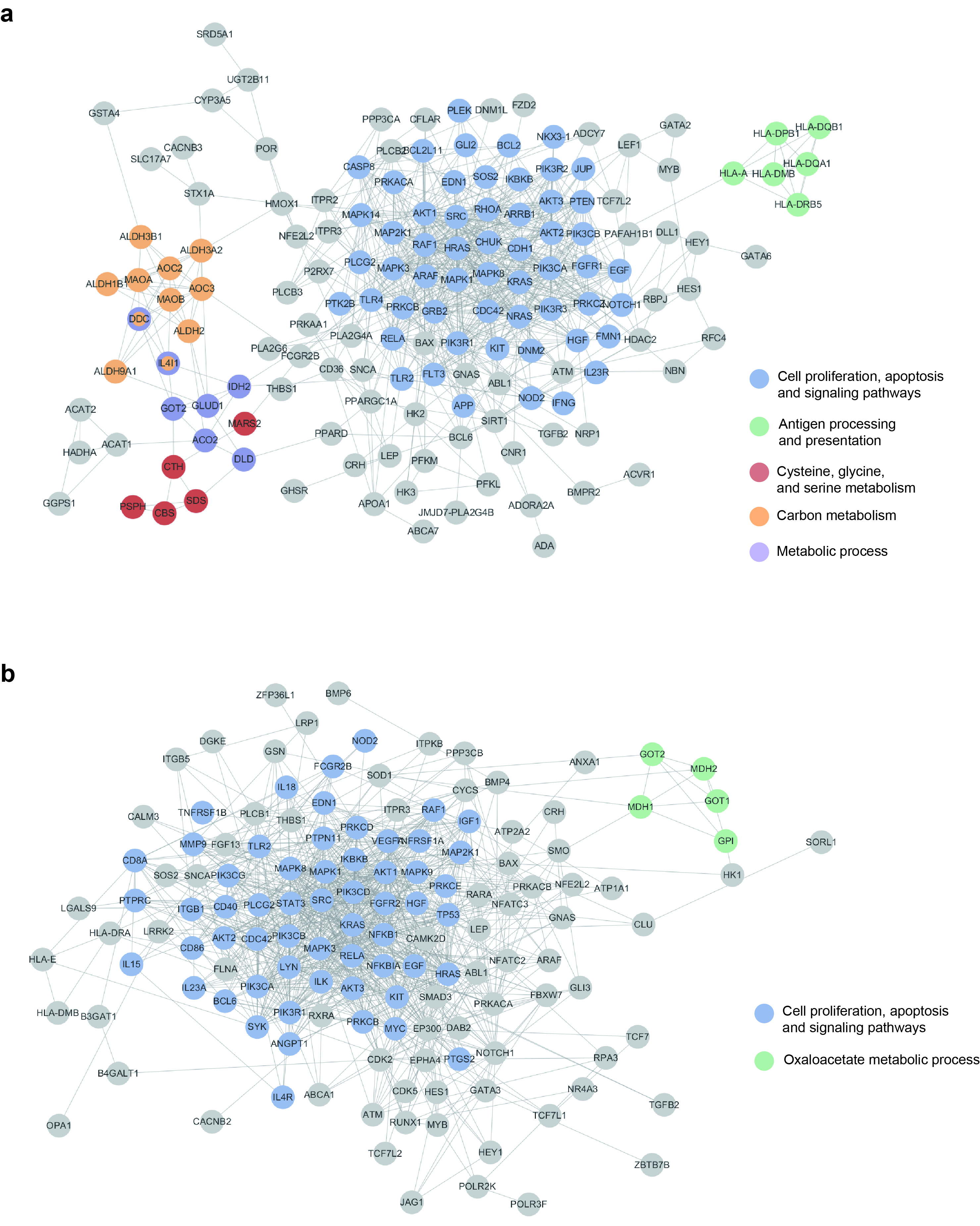}\par
\caption{Mapping highly contributing genes to the PPI network. Colored nodes are sub-networks clustered using the ClusterOne algorithm. Gene nodes can be mapped to more than one sub-networks. (a) The blood sub-network. (b) The brain sub-network.} 
\label{fig:fig3}
\end{figure}

\subsection{Finding sub-networks of highly contributed genes}
Using a protein interaction network obtained from the STRING database, we examined genes with high importance scores. We selected 1\% genes with the highest importance scores from the four models. We integrated genes selected in PINNet (GO) and PINNet (KEGG) models for the blood and brain, separately, and obtained 170 and 167 genes in blood and the brain, respectively. The highly contributed genes were mapped onto the STRING database (version 11.0) \cite{szklarczyk2019string}, where interactions larger than a confidence score of 0.7 were used. Cytoscape\cite{shannon2003cytoscape} was used to configure the PPI network. As shown in Fig. 3, 152 and 138 genes in the blood and brain were mapped in the largest network, respectively. We also used ClusterOne\cite{nepusz2012detecting} to screen PPI subnetwork modules. ClusterOne identifies overlapping functional modules by detecting dense regions in the protein interactome network. The parameters were set to $p$-values $<$ 0.05, the minimum size = 5, and an edge weight as a confidence score in the STRING database. If genes of two subnetworks were overlapped by more than 70\%, we merged them into a large module. Then, modules were selected if more than one biological term was enriched with GO biological process (BP) and KEGG using DAVID \cite{sherman2009systematic}(Bonferroni corrected $p$-values $<$ 0.05). Figure 3 shows blue, green, red, orange, and purple modules in the blood and blue and green modules in the brain. In the blood, the blue module was significantly enriched in 218 terms related to proliferation and several signaling pathways, including Ras, FOXO, and PI3K-Akt signaling pathways. The green module was enriched with antigen processing and presentation terms, especially the MHC class II protein. In the red module, the serine family, comprising cysteine, glycine, and serine metabolism-related terms, was enriched. The orange module was enriched with several carbon metabolism pathways. The purple module was related to several metabolic processes, including glycolysis, ethanol catabolic process, and aldehydecatabolic process. In the brain, the blue module was similar to the blue module in blood. It was enriched with 350 terms related to proliferation and several signaling pathways, including TNF, FOXO, and PI3K-Akt signaling pathways. The green module contained 29 enriched terms related to the oxaloacetate metabolic process.

\subsection{Comparisons of contributions of genes on blood and brain dataset}
We also identified highly contributed genes that show differences between blood and brain in AD prediction. First, we compared models based on the same pathway database (GO, KEGG) in different tissues with a two-sided Wilcoxon test. From the results of 10-fold cross-validation, absolute SHAP values with z-score normalization were used for the test, and the p-values were corrected by Benjamini–Hochberg method (BH). We obtained 24 and 291 significantly different genes between blood and brain models (blood (GO) vs. brain (GO) and blood (KEGG) vs. brain (KEGG), respectively) with the adjusted $p$-values (BH) $<$ 0.05. Next, we checked highly contributed genes (top 1\%) that showed a clear distinction between the tissues. As a result, in the brain, CLU, PTPN11, ITPKB, PRKACB, IGF1, and MTHFD2 were more important compared to blood. CLU is related to several pathological states of AD, including a mediator of A$\beta$ toxicity\cite{foster2019clusterin}. PTPN11 (also known as SHP2) interacts with tau in Alzheimer's disease brain\cite{kim2019tau}. Overexpression of ITPKB induces the increase of A$\beta$40 and tau hyperphosphorylation in a mouse model of familial Alzheimer's disease\cite{stygelbout2014inositol}. PRKACB is involved in the elevation of A$\beta$ and tau hyperphosphorylation levels\cite{wang2019microrna}. Reduced IGF-1 signaling in the AD mouse model is attributed to a decline in neuronal loss and behavioral impairment\cite{Cohen2009reduced}. MTHFD2 is differentially expressed in AD posterior cingulate astrocytes\cite{sekar2015alzheimer}. In the case of blood, although two genes were significantly important compared  to brain, there is no known evidence to explain the relevance of these genes and AD in blood.

\subsection{Prediction performance of genes with high importance scores}
We further investigated whether genes selected using the importance scores in PINNet can be relevant features for predicting AD. First, we divided all samples into ten sets; eight sets were used for training, a set for selecting genes (named a feature selection set), and a remaining set for testing. Second, after training PINNet using the training set, we selected 892 (10\%) genes with the highest importance scores in the feature selection set. Third, a multilayer perceptron (MLP) model was trained with the 892 selected feature genes, and then the remaining test set was used to evaluate the AD prediction performance. We repeated this process ten times. The MLP model consists of four layers, and the numbers of nodes in each layer are \{$n$, $n/2$, $n/4$, $2$\}, where $n$ is the number of input genes. We compared the performance of 10\% of selected genes using PINNet with the same number of randomly selected genes and all genes for predicting AD samples. Genes selected using PINNet outperformed the random gene set in blood models (Table \ref{tab:table3}), suggesting that only 10\% of the genes selected from PINNet can represent the whole blood transcriptome dataset.

\begin{table}[ht]
\centering
\begin{tabular}{|c|c|c|cc|}
\hline
\multirow{2}{*}{} & \multirow{2}{*}{All (8922)} & \multirow{2}{*}{Random (892)} & \multicolumn{2}{c|}{Selected Features (892)}    \\ \cline{4-5} 
                  &                             &                               & \multicolumn{1}{c|}{PINNet (GO)} & PINNet (KEGG) \\ \hline
Blood             & 0.6092$\pm$0.0762               & 0.5745$\pm$0.0875                 & \multicolumn{1}{c|}{0.5938$\pm$0.0776}  & 0.6031$\pm$0.0598    \\ \hline
Brain             & 0.9764$\pm$0.0258               & 0.9750$\pm$0.0440                 & \multicolumn{1}{c|}{0.9720$\pm$0.0343}  & 0.9743$\pm$0.0314    \\ \hline
\end{tabular}
\caption{Performance comparisons (AUC) using all genes, random genes and selected genes from PINNet. AUC, area under the curve; DNN, deep neural network}
\label{tab:table3}
\end{table}

\section{Discussion}
In this study, we developed PINNet, a model with higher interpretability than a black box neural network with pathway information as prior knowledge. When designing a model, relationships between pathways and genes were represented as weights between the input layer and the pathway layer. 

We further examined the importance score of pathway nodes in the pathway layer using the same ten models that analyzed the importance scores of input genes. The maximum values of importance scores for input gene nodes and pathway nodes were 11.904 and 1.512, respectively, and the variances were 0.697 and 0.321, respectively. Since the importance scores were defined as the average of $z$-scores of absolute mean SHAP values, this result means that there were relatively few pathway nodes having high importance scores across all ten models. Furthermore, there was no significant difference in importance scores of pathway nodes between blood and brain models. When the Wilcoxon test was performed, adjusted $p$-values (BH) were $>$ 0.05 for all pathways.

Although the average importance scores of pathway nodes were relatively smaller than compared to those of gene nodes, pathways with high importance scores were related to AD. The characteristics of the highly ranked pathways in the blood were related to the glycosylphosphatidylinositol (GPI) anchor-related pathway and immune response. The GPI anchor metabolic process was the highest-ranked GO BP, and similar KEGG pathways, such as glycerophospholipid metabolism and GPI-anchor biosynthesis, were also highly ranked  among pathway nodes of blood (KEGG). In addition, several immune-related terms have been ranked top in blood. In recent studies, inflammation has been identified as an important contributor to AD pathology. Peripheral immune cells, such as T cells, are activated and infiltrate the inflamed brain area through the damaged blood-brain barrier (BBB) and accumulate in the AD brain \cite{town2005t, togo2002occurrence}. Moreover, disrupted BBB may allow complement proteins to reach the brain from the plasma, and cell damage and death can lead to complement activation of neurons and oligodendrocytes, which activate complement and lead to dysregulation \cite{morgan2018complement}.

In the prediction of the brain dataset, the GO BP term with the highest importance score is the regulation of filopodium assembly. Filopodia is the protrusion at the end of the neuron and is related to neural plasticity\cite{ozcan2017filopodia}. AD pathology shows a correlation with filopodia density\cite{boros2019dendritic}. In KEGG, neurodegenerative diseases such as Parkinson's disease, Alzheimer's disease, and prion disease showed high importance scores. SNARE interactions in vesicular transport ranked 9th on pathway nodes of the brain (KEGG), and it is associated with neurotransmitter release \cite{han2017multifaceted}. In particular, A$\beta$ oligomers interfere with SNARE-medicated vesicle fusion, which may cause synaptic dysfunctions \cite{yang2015amyloid}. Similar to blood models having pathways related to immune response, many immune-related pathways were also important in the brain models. We suggest that immune-related genes have the potential to advance the understanding of AD in terms of systemic disease in future studies.

\section{Conclusion}
We proposed a pathway information-based neural network model for the AD prediction. PINNet applied pathway-level of prior biological knowledge in constructing connections between genes and the AD status. We improved classification accuracies of AD and CN in blood and brain using PINNet, and the interpretation of the trained model  revealed biological systems related to AD. Additionally, PINNet would be a promising model for various disease prediction tasks due to its possibility of model interpretability. 

\section{Methods}
\subsection{Preprocessing of data}
Gene expression datasets from the prefrontal cortex brain and blood were obtained from GSE33000 \cite{narayanan2014common} and ADNI (adni.loni.usc.edu)\cite{petersen2010alzheimer}, respectively. Brain and blood gene expression data were generated using the Rosetta/Merck Human 44k 1.1 microarray \cite{narayanan2014common} and the Affymetrix Human Genome U 219 array \cite{petersen2010alzheimer}, respectively. We used 113 AD and 244 CN samples for the blood dataset after excluding mild cognitive impairment (MCI) samples. For the brain dataset, we used 310 AD and 157 CN samples after excluding Huntington's disease samples. The blood and brain datasets consist of 49,386 and 39,328 probes, respectively. The mean imputation was performed on the remaining missing data values in each dataset. We removed the 30\% of probes with a low interquartile range (IQR) across samples. To compare gene expression across different platforms, we mapped probe IDs to Entrez IDs using the biomaRt package \cite{10}. The probe with the maximum IQR value was selected for multiple probes annotated with the same gene. As a result, 13,666 and 12,319 genes remained in the brain and blood datasets, respectively. A total of 8,922 genes were common between the brain and blood datasets, and they were used in this study. 

Pathway information was compiled from the GO BP \cite{12} and KEGG\cite{11} databases obtained from the Molecular Signature Database (MSigDB)\cite{13}. A total of 168 KEGG pathways and 4,026 GO BPs were selected. They include ten or more genes among the 8,922 genes used in this study. Using each of the two databases, a pathway information matrix $M \in \mathbb{R}^{m \times n}$ is constructed, which is a sparse matrix representing the relationship between pathways and genes (Eq. (1)), where $m$ is the number of pathways and $n$ is the number of genes.

\begin{ceqn}
\begin{align}
\label{eq:eq1}
M_{ij} = \begin{cases}0 & j_{gene}\notin i_{pathway} \\1 & j_{gene}\in i_{pathway}\end{cases}
\end{align}
\end{ceqn}

\subsection{Structure and training of PINNet}
PINNet is a four-layer deep neural network with biological pathway information as prior knowledge for AD classification. The PINNet comprises four layers: an input layer, a pathway layer, a hidden layer, and an output layer, as shown in Figure 1. Gene expression values of a sample $g\in \mathbb{R}^n$ are allocated to input nodes. Thus, the number of input nodes ($n$) is the same as the number of genes ($n = 8,922$) in the dataset. The pathway layer consists of fully connected nodes $f$ and pathway nodes $p$. Pathway nodes $p$ represent relations between genes and pathways, and fully connected nodes $f$ represent features for all genes regardless that they are included in the pathway. The number of nodes in $p\in \mathbb{R}^m$ was the same as the number of pathways used. For instance, $m$ is 7,470 when using the GO database and 186 when using the KEGG database. In order to represent the relationship between the gene and the pathway, we obtain pathway nodes $p$ as 
\begin{ceqn}
\begin{align}
\label{eq:eq3}
p = tanh(((W_p \circ M) \times g) \circ u),
\end{align}
\end{ceqn} where $W_p$ is weight matrices. The $W_p \in \mathbb{R}^{m \times n}$ is masked by the pathway information matrix $M \in \mathbb{R}^{m \times n}$ with element-wise multiplication. Masking weights ($W_p \circ M$) produce a sparse network by weighting between a particular pathway and genes. Genes not included in the pathway are set to zero. Note that non-zero weights of $W_p$ are updated by backpropagation. $u\in \mathbb{R}^m$ is a normalization vector to prevent the values of pathway nodes are affected by the number of genes in the pathway and is defined as follows:
\begin{ceqn}
\begin{align}
\label{eq:eq2}
u_i = \frac{1}{\sqrt{\sum_{j}^{n}M_{ij}}}
\end{align}
\end{ceqn} 
For the fully connected nodes $f$, $n$ input nodes are directly connected to the $f$ through a linear layer as follows: 

\begin{ceqn}
\begin{align}
\label{eq:eq4}
f = tanh(W_f g)
\end{align}
\end{ceqn} where $W_f$ is a weight matrix. $p$ and $f$ are concatenated after layer normalization ($[LayerNorm(p);LayerNorm(f)]$), and it passes through the hidden layer and the output layer. Then, output $o  \in \mathbb{R}$ is obtained. Tanh is used for the nonlinear transforming function, and batch normalization is utilized to decrease the effect of internal covariate shift induced by various weight scales. Dropout ($\alpha$ = 0.3) prevents overfitting of the model. We performed the training process by minimizing a cross entropy loss. 

The training dataset was oversampled using SMOTE\cite{14} to avoid data imbalance. SMOTE oversampling creates synthetic instances and generates a new instance in line, a segment of the randomly selected k-nearest neighborhood of the minority class. To train PINNet, we initialized all weights with the Xavier initialization. We optimized the loss using the ADAM optimizer, the ReduceLROnPlateau scheduler, and a stochastic gradient descent algorithm, with a mini-batch size of 64. Note that the same settings are applied to both PINNet and the neural network model used for the comparison. The model was implemented using Python (version 3.7.4, \url{https://www.python.org/}) and Pytorch (version 1.8.1, \url{https://www.pytorch.org/}) on GeForce TITAN X GPUs.

\subsection{Interpretation of PINNet and calculation of importance scores}
We interpreted the model to identify highly contributing genes and pathways using the SHAP values. The SHAP values for each model were computed using the DeepExplainer SHAP package in Python, which is based on Deep SHAPv\cite{lundberg2017unified}. DeepExplainer calculates the feature importance of a given input to predict the deep neural network. The SHAP value is close to 0 if the feature does not affect the model’s expectation.
Deep SHAP explains the difference between output $y$ and reference output $\bar{y}$ by the summation of the difference between inputs $x$ and reference inputs $\bar{x}$. Reference output $\bar{y}$ are determined by executing forward passes in the network under reference inputs $\bar{x}$. Let $\triangle x_i$ be the difference between input $x_i$ and reference input $\bar{x_i}$ and $C_{\triangle x_i}^{(l)}$ be the attribution of node $i$ in the $l$th layer. Then, the difference between output $y$ and $\bar{y}$ can be explained by the summation of attribution scores of $x_i$ in the input layer $l$: 
\begin{ceqn}
\begin{align}
\label{eq:eq5}
\sum_{i}^{} C_{\triangle x_i}^{(l)} = y-\bar{y}
\end{align}

\end{ceqn} 
Reference inputs and outputs can be chosen for a given problem, and we set the reference to a training set with a balanced label distribution, where the reference output is the average of the outputs of the training set. The attribution score calculated from Deep SHAP can be adapted to an approximation of SHAP values. To obtain the contribution of the genes in the input layer, $x_i$ in Eq. (\ref{eq:eq5}) is gene expression values in a test set. Furthermore, to compute the contribution of pathway nodes in the pathway layer, the values of pathway nodes $p$ in Eq. (\ref{eq:eq3}) are used as $x_i$ in Eq. (\ref{eq:eq5}). For each pathway and brain and blood datasets, ten models are constructed by 10-fold cross-validation. Absolute mean SHAP values are normalized as $z$-scores, and the $z$-scores are averaged for the ten models. We defined it as an importance score. Importance scores are used to evaluate the contribution of the genes and pathways. 

\section{Availability of data and materials}
Our source codes and model are available at \url{https://github.com/DMCB-GIST/PINNet}. The gene expression datasets from the prefrontal cortex brain in this study are publicly available through the NCBI GEO database with accession number GSE33000. The blood gene expression datasets are available from the ADNI database (http://adni.loni.usc.edu). Gene sets of GO and KEGG were obtained from the MSigDB (https://www.gsea-msigdb.org/gsea/msigdb).

\bibliographystyle{unsrt}  
\bibliography{references} 

\section*{Acknowledgements}
Data collection and sharing for this project was funded by the Alzheimer's Disease Neuroimaging Initiative (ADNI) (National Institutes of Health Grant U01 AG024904) and DOD ADNI (Department of Defense award number W81XWH-12-2-0012). 
ADNI is funded by the National Institute on Aging, the National Institute of Biomedical Imaging and Bioengineering, and through generous contributions from AbbVie, Alzheimer's Association; Alzheimer's Drug Discovery Foundation; Araclon Biotech; BioClinica, Inc.; Biogen; Bristol-Myers Squibb Company; CereSpir, Inc.; Cogstate; Eisai Inc.; Elan Pharmaceuticals, Inc.; Eli Lilly and Company; EuroImmun; F. Hoffmann-La Roche Ltd and its affiliated company Genentech, Inc.; Fujirebio; GE Healthcare; IXICO Ltd.; Janssen Alzheimer Immunotherapy Research {\&} Development, LLC.; Johnson {\&} Johnson Pharmaceutical Research {\&} Development LLC.; Lumosity; Lundbeck; Merck {\&} Co., Inc.;Meso Scale Diagnostics, LLC.; NeuroRx Research; Neurotrack Technologies; Novartis Pharmaceuticals Corporation; Pfizer Inc.; Piramal Imaging; Servier; Takeda Pharmaceutical Company; and Transition Therapeutics. The Canadian Institute of Health Research is providing funds to support the ADNI clinical sites in Canada. 
Private sector contributions are facilitated by the Foundation for the National Institutes of Health (www.fnih.org). The grantee organization is the Northern California Institute for Research and Education, and the study was coordinated by the Alzheimer's Therapeutic Research Institute at the University of Southern California. ADNI data were disseminated by the Laboratory for Neuro Imaging at the University of Southern California.

\section*{Author contributions statement}
HL initiated and supervised the study. YK developed the method and performed the experiments. All authors participated in writing the manuscript. Both authors have read and approved the manuscript.


\end{document}